\documentclass[10pt, preprint]{aastex}
\usepackage{amsmath,amssymb,lmodern,mathrsfs, ulem, abstract}
\usepackage[caption=false]{subfig}

\newcommand{\grad}{\mbox{\boldmath$\nabla$}}
\title{Peristaltic Pumping near Post-CME Supra-Arcade Current Sheets}
\author{Roger B. Scott} 
\author{Dana W. Longcope}
\author{David E. McKenzie}

\begin{document}

\maketitle

\begin{abstract}
Measurements of temperature and density near supra-arcade current sheets suggest that plasma on unreconnected field lines may experience some degree of ``pre-heating'' and ``pre-densification'' prior to their reconnection. Models of patchy reconnection allow for heating and acceleration of plasma along reconnected field lines but do not offer a mechanism for transport of thermal energy across field lines. Here we present a model in which a reconnected flux tube retracts, deforming the surrounding layer of unreconnected field. The deformation creates constrictions that act as peristaltic pumps, driving plasma flow along affected field lines. Under certain circumstances these flows lead to shocks that can extend far out into the unreconnected field, altering the plasma properties in the affected region. These findings have direct implications for observations in the solar corona, particularly in regard to such phenomena as high temperatures near current sheets in eruptive solar flares and wakes seen in the form of descending regions of density depletion or supra-arcade downflows.
\end{abstract}


\section{Introduction}

Since the development of X-ray and EUV solar imaging, observations of evolving arcade structures have become a ubiquitous signature of magnetic reconnection in solar flares. Many of these structures also exhibit vertical fans with highly emissive coronal plasma and what is presumed to be a nearly vertical magnetic field rising above the apex of the arcade \citep{Svestka_1998, McKenzie_1999, Webb_2003}. This picture is consistent with the standard flare model in which a current sheet separates antiparallel layers of magnetic field between an arcade of reconnected flux and a rising coronal mass ejection \citep{Cliver_2002}. But while the general properties of these structures are well established, the mechanism responsible for increased emission from plasma in the supra-arcade fan remains unclear \citep{Seaton_2009, Ko_2010, Reeves_2010}. 

One possibility is that the emitting plasma is within the current sheet itself and that its temperature has been increased as a result of ohmic heating. This explanation relies on the assumption that the line-of-sight depth of the current sheet is large enough to allow for a non-negligible emission measure. In cases where the current sheet is observed edge-on (or nearly so), the line-of-sight depth can easily exceed $10^5$ km.  These edge-on observations \citep[e.g.,][]{Ciaravella_2008, Savage_2010} also enable upper limits to be placed on the thickness of the current sheet: the measurements indicate thicknesses of no more than 5--50 $\times 10^3$ km.  Conversely, \cite{Tucker_1973} used theoretical arguments and estimated that post-CME current sheets should have a thickness of roughly $10^{-1}$ km.  Such thickness estimates become crucial in cases where the current sheets are observed face-on, as in \cite{Svestka_1998}, \cite{McKenzie_1999}, \cite{Innes_2003}, \cite{Warren_2011}, and \cite{McKenzie_2013}.

An alternative explanation is that the emission comes not from within the current sheet itself but rather from a {\it thermal halo} that surrounds the current sheet. The thermal halo could be orders of magnitude thicker than the current sheet and thus provide sufficient line-of-sight depth for observed emission. However, this scenario requires some mechanism for increasing the local plasma density above that of the surrounding corona. Chromospheric evaporation is a likely candidate for this pre-densification, but one must still justify the timeliness of the evaporation, which is usually attributed to thermal conduction into the chromosphere \citep{Cargill_1995}. 

Reconnection within the current sheet is a likely source of energy both for the heating of plasma and for evaporation-driven pre-densification because it efficiently converts magnetic free energy into thermal and kinetic energy \citep{Guidoni_2010, Priest_1999}. But while reconnection may provide sufficient energy to heat the surrounding plasma, thermal conductivity transverse to the magnetic field is very weak \citep{Choudhuri}. Even if we assume that there exists a well of thermal energy in the reconnected field, it remains unclear what mechanism could be responsible for transporting the energy across field lines. And while radiative transfer is not limited by thermal conduction it is also far too weak given the low optical depth that is typical of the corona. Nonlinear mode-coupling could play a role if the reconnection event somehow excited magnetosonic waves that then dissipated energy in the surrounding plasma.

Recent observations in EUV \citep{Savage_2012, Warren_2011, Savage_2011} have resolved what appear to be magnetic loops that descend through the supra-arcade fan. The loops seem to form wake-like structures that appear as density depletions or voids in the surrounding plasma. The nature of these voids was studied by \cite{Verwichte_2005}, who characterized the apparent wave motion of their edges. They found that the boundary between the low density voids and the surrounding plasma exhibited transverse oscillating wavelets that propagated sunward at speeds in the range of 50 km s$^{-1}$ to 500 km s$^{-1}$. \cite{Costa_2009} simulated the formation of these dark lanes from an initial pressure perturbation. They found that the lanes could be interpreted as an interference pattern resulting from the reflection of magnetosonic shocks and rarefaction waves. More recently, \cite{Cassak_2013} simulated the formation of dark lanes as ``flow channels carved by sunward-directed outflow jets from reconnection.'' The applicability of this last interpretation, which places the voids below the arcade itself, must be carefully considered when placed in the context of observations of supra-arcade features.

\begin{figure}[ht]
\begin{center}
\subfloat[]{\label{cartoon1.sub}\includegraphics[width = 0.35\textwidth]{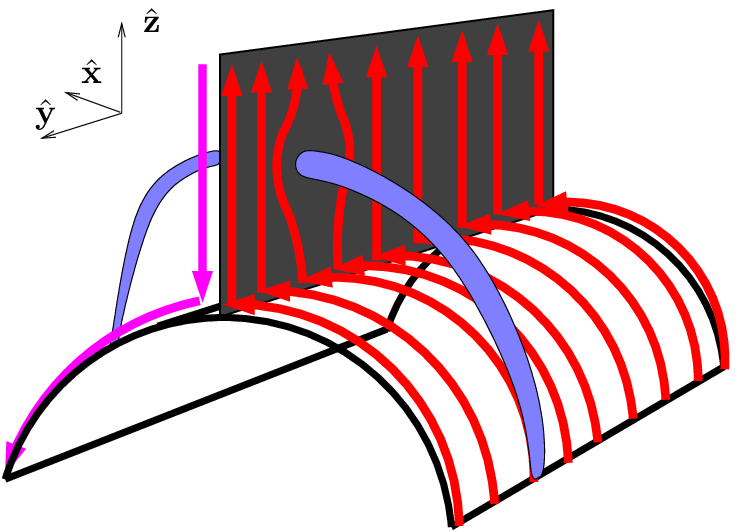}}
\subfloat[]{\label{cartoon2.sub}\includegraphics[width = 0.35\textwidth]{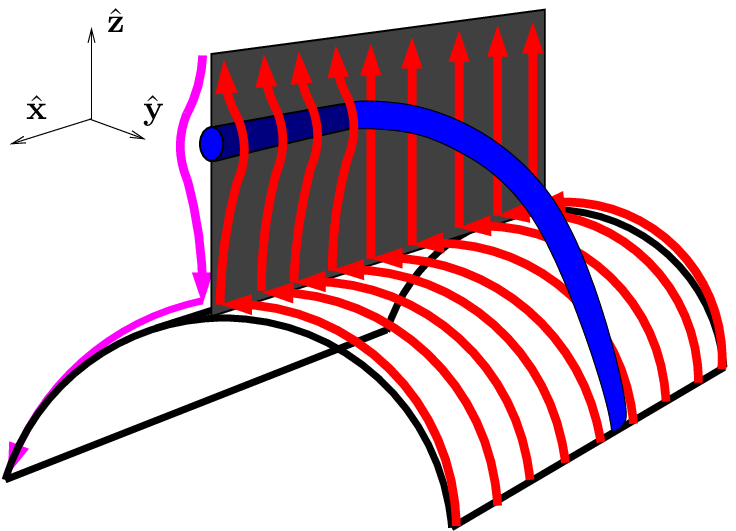}}
\caption{\label{cartoon.fig}  (\ref{cartoon1.sub}) A reconnected flux tube piercing normally through the current sheet as in \cite{Savage_2012}.  (\ref{cartoon2.sub}) A reconnected flux tube embedded within the current sheet as in \cite{Linton_2006}. Taking $\hat{\bf z}$ to point vertically away from the limb with $\hat{\bf x}$ pointing along the reconnected flux tube, $\hat{\bf y}$ is either normal to or in the plane of the current sheet, depending on the configuration. }
\end{center}
\end{figure}

Another possibility is that these features are the result of patchy reconnection in which flux tubes retract toward the arcade under the influence of magnetic tension and are drawn through the surrounding, unreconnected field as depicted in Figure \ref{cartoon.fig}.  Previous authors have modeled the dynamics of the retracting flux tube \citep{Guidoni_2010, Longcope_2009, Linton_2006}, but have not yet considered its effect on the surrounding, unreconnected flux.  \cite{Cargill_1996} modeled the interaction of a magnetic cloud and the surrounding magnetic field, but their work focused on the high-$\beta$ regime ($\beta = 8 \pi p / B^2 \gg 1$). For our analysis we will consider the consequences of an extremely low-$\beta$ scenario in which the magnetic field dominates all other energy contributions. 

Our focus will be to consider how the plasma and unreconnected flux that surround the current sheet behave in response to a reconnection event. Toward this end we assume that a localized reconnection event has already occurred within a supra-arcade current sheet and has created a bundle of newly closed magnetic field lines, a flux tube, which retracts through the current sheet \citep{Linton_2006, Longcope_2009} as depicted in Figure \ref{cartoon.fig}. The retracting flux tube is a prescribed element whose radius and motion are parameters of the model. The primary effect we consider is the deformation it creates in the surrounding field.  The deflection of a given field line is bounded by the radius of the retracting tube and is smaller farther away.  Due to this smallness the deformation is typically dismissed as a minor effect, though \citet{Linton_2006} did consider the possibility that the work required to the deform the external field might contribute to a drag force on the retracting tube. 

The retracting flux tube could have two possible orientations relative to the current sheet.  The theoretical work of \citet{Linton_2006} assumes that a section of the tube lies within the plane of the current sheet, as shown Figure \ref{cartoon2.sub}.  On the other hand imaging observations have been interpreted assuming the flux tube pierces the current sheet normally, as in Figure \ref{cartoon1.sub} \citep{McKenzie_2000, Savage_2012}.  Our modeling will be applicable to either scenario since both create identical deformations in the surrounding field.  We hereafter focus on the surrounding field which is roughly vertical, and refer to the retracting flux as an {\it intrusion}. 

In the present work we show that the deformation takes the form of a constriction, which moves downward through the surrounding field at the same speed as the retracting flux tube.  Observations clearly show this speed to be some fraction of the local Alfv\'en speed \citep{Savage_2011}, and often in excess of the local sound speed.  We observe that the moving constriction behaves as a peristaltic pump, resulting in field-aligned plasma flows, which we dub {\it peristaltic flows}. We show below that there are regimes in which these flows lead to slow magnetosonic shocks that develop in the surrounding field. These manifest in our model as hydrodynamic shocks and rarefaction waves, which travel along the field at speeds comparable to the sound speed. The existence of such features leads to several dramatic effects, including significant heating and changes to the density and emission measure of plasma in the unreconnected field.


\section{The Model}

The model that we present here treats the unreconnected flux as current-free field along which plasma is constrained to move. We begin by determining the magnetic field subject to the influence of an intruding, reconnected flux tube. We assume that $\beta$ is extremely small so that the magnetic field may be determined independent of the plasma. This dictates both the plasma flow trajectory and the cross section of parallel flow. Steady solutions are then found for plasma flow along each field line. Points where the flow is ill-defined are avoided through the introduction of rarefaction waves and acoustic shocks, which are a limiting form of slow magnetosonic shocks in very low $\beta$. The result is a piecewise continuous adiabatic series of solutions that evolve in time as the fluid jumps propagate. The 2D behavior is ultimately found through interpolating between solutions along representative field lines. 

Our analysis will invoke two distinct reference frames. The {\it limb-frame} is stationary with respect to the solar surface and in this frame the undisturbed plasma is at rest. Alternatively, in the {\it comoving frame} it is the descending intrusion that is at rest and the plasma is taken to be rising uniformally at large distances from the intrusion. It is in the comoving frame that the magnetic field is most easily determined because in this frame the boundary conditions are steady in time and therefore, so too is the field.


\subsection{Deformed Potential Field}

In the comoving frame the unreconnected magnetic field is a sum of the original magnetic field prior to distortion $({\bf B}_0)$ and a second field $({\bf B}^\prime)$ representing the influence of the intruding flux tube
\begin{equation}
{\bf B} = {\bf B}_0 + {\bf B}^\prime.
\end{equation}
Since the reconnected and unreconnected fields are topologically distinct we will impose the simplifying constraint that no unreconnected field lines may intersect the surface of the reconnected flux tube. Furthermore, as information about the reconnection event cannot have influenced the field at arbitrarily large distances, ${\bf B}$ must reduce to ${\bf B}_0$ far from the flux tube, so ${\bf B}^\prime$ must vanish there. 

\begin{figure}[ht!]
\begin{center}
\includegraphics[width = 0.4\textwidth]{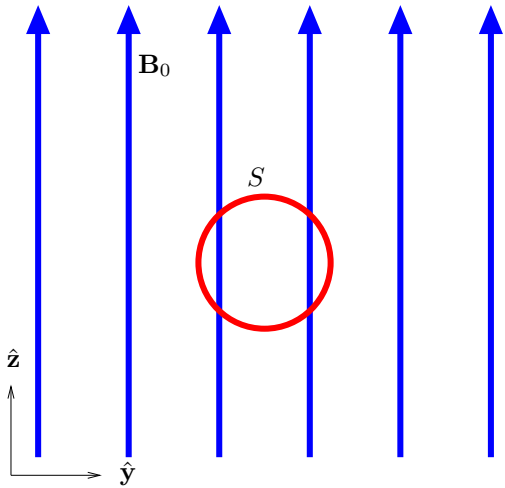}
\caption{\label{setup.fig} An initially uniform field (${\bf B}_0$) is altered by the intruding surface ($S$) with ${\bf B}^\prime$ introduced so that the net field ($ {\bf B} = {\bf B}_0 + {\bf B}^\prime $) satisfies the appropriate boundary conditions on $S$. }
\end{center}
\end{figure}

We take the background field to be uniform and vertical (${\bf B}_0 = \pm \hat{\bf z} B_0$) while the reconnected flux defines a uniform cylinder ($S$), centered at the origin, with radius $R$ and symmetry axis pointing in the $\hat{\bf x}$ direction.\footnote{This assumes that the radius of curvature of the reconnected flux is large compared to the embedded length within the fan.} ${\bf B}^\prime$ depends only on $y$ and $z$. The total field is assumed to be potential with boundary conditions given by 
\begin{equation}
\left . \hat{\bf z} \times {\bf B} \right |_{r \rightarrow \infty} = \left . \hat{\bf r} \cdot {\bf B} \right |_{r \in S} = 0.
\end{equation}
The first constraint ensures that the magnetic field is unaffected far from $S$ while the latter ensures that no field lines intersect $S$.

The potential magnetic field, constrained by these boundary conditions, may be determined in terms of a flux function such that
\begin{equation}
{\bf B} = \hat{\bf x} \times B_0\grad f = - B_0 \grad \times (\hat{\bf x} f), 
\end{equation}
where $\hat{\bf x}$, being the axis of symmetry of the intruding flux, is an ignorable direction. In the far field, setting $f \rightarrow y$ ensures that ${\bf B}\rightarrow B_0 \hat{\bf z}$, while on the surface of the feature $\hat{\bf r} \cdot {\bf B} = 0$ so $\left .\partial_\theta f \right |_S= 0$.  Thus, in terms of $f$ the boundary conditions become 
\begin{equation}
\left . \grad f \times \hat{\bf y} \right |_{r \rightarrow \infty} = 0
\end{equation}
and
\begin{equation}
\left . \grad f \times \hat{\bf r} \right |_{r \rightarrow R} = 0. 
\end{equation}
And, since $\grad \times {\bf B} = 0$, $f$ satisfies Laplace's equation;
\begin{equation}
\nabla^2 f = 0.
\end{equation} 

\begin{figure}[ht!]
\begin{center}
\includegraphics[width = 0.5\textwidth]{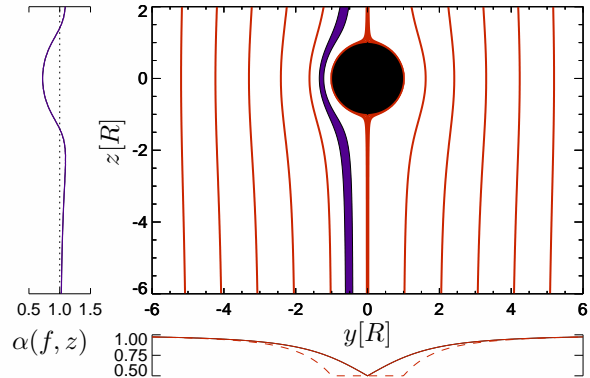}
\caption{\label{field.fig}The potential magnetic field is deformed by the expansion of the origin onto a cylindrical surface of radius R. Field lines are deflected around the intrusion. The inverse normalized field strength ($\alpha$) is shown on the left panel for the $f =  - R/2$ field line, parameterized in $z$. The associated flux tube is traced in purple on the main panel. In the bottom panel the minimum inverse normalized field strength ($ \alpha_{z = 0}$) is plotted first as a function of $f$ (solid red) and then as a function of $y$ (dashed red). }
\end{center}
\end{figure}

With these conditions and the choice that $f$ be symmetric in $y$, the flux function is uniquely specified as
\begin{equation}
f = \sin(\theta) \left ( \frac{R^2}{r} - r \right ) = y \left ( 1 - \frac{R^2}{y^2 + z^2} \right ) \label{f.eq},
\end{equation}
with $\theta$ measured from $-\hat{\bf z}$. 
Since the magnetic field is everywhere orthogonal to the gradient of $f$, contours of $f$ are themselves field lines, denoted ${\bf X}_f$, which can be parameterized by solving for $z$ in terms of $f$ and $y$ so that
\begin{equation}\label{line_param.eq}
z_f(y)^2 = y \frac{y^2 - f y - R^2}{f - y},
\end{equation}
where, for a given $f$, the $y$ coordinate along the field line is bounded by
\begin{equation}
|f| < |y| < \sqrt{f^2/4 + 1} + |f|/2.
\end{equation}
Figure \ref{field.fig} shows a contour plot of $f(y,z)$ that traces a representative set of field lines. For each field line, $y \rightarrow f$ as $|z| \rightarrow \infty$. Values of $f$ are therefore the lateral positions of the field lines in the far field. The deflection of each field line is largest abreast of the intrusion, where $z = 0$. For $|f| \gg R$ this deflection goes to zero while for the most strongly deflected field line ($f = 0$) the deflection is $|\Delta y| =  R$. 

Ultimately, we will be interested in the parameterized cross section of an arbitrary, unreconnected flux tube. From $\grad \cdot {\bf B} = 0$ it follows that the cross section scales inversely with the field strength. Let $\alpha({\bf X}_f)$ be the inverse of the dimensionless field strength of an infinitesimal flux tube, normalized to unity in the far field and parameterized along an arbitrary field line, ${\bf X}_f$, so that $1/\alpha \equiv |{\bf B}| / B_o$. In terms of $f$
\begin{align}\label{2dcross.eq}
\frac{1}{\alpha^2} = & (\partial_{y} f)^2 + (\partial_{z} f)^2 \\ 
 = & \frac{2 R^2 (y^2 - z^2) + R^4}{(y^2 + z^2)^2} + 1,
\end{align}
which, after substituting in for $z_f(y)$, can be expressed as
\begin{equation}\label{cross.eq}
\alpha^2 =  \frac{y^2 R^2}{f^2 R^2 + 4 y^2 \left ( f - y \right )^2}.
\end{equation}
Holding $f$ fixed, we define $\alpha_f(y)$ to be the cross section of an infinitesimal flux tube centered on a field line ${\bf X}_f$, parameterized by the lateral deflection of the field line.

$\alpha_f$ achieves a minimum value at $z = 0$, where the field line passes abreast of the intrusion. This location is referred to as the throat of the flux tube and has a cross section of 
\begin{equation}\label{alpha_min.eq}
\alpha_{min}(f) = \frac{1}{2} \frac{\sqrt{f^2 + 4 R^2} + f}{\sqrt{f^2 + 4 R^2}},
\end{equation}
which is necessarily less than one. Moving away from the throat along the field line the flux tube expands until it reaches a maximum value, which is necessarily greater than one, and then slowly contracts toward unity as $|z|\rightarrow \infty$. In general, field lines that pass close to the intrusion have the smallest minimum cross section and greatest overall variability while for large $|f|$ values $\alpha$ is nearly uniform along $z$. The $f = 0$ field line is both the most and least constricted with a cross section that diverges at $y = 0, z = \pm R$ and achieves the global minimum of $\alpha_{min}(f = 0) = \min[\alpha(y,z)] = 0.5$ at its throat.


\subsection{Peristaltic Flow}\label{pflow.sec}

Under the assumption of ideal magnetic induction, as a fluid element moves it must remain on the same field line and its cross section for flow parallel to the field must be the same as that of the associated flux tube. Since the magnetic field is stationary with respect to the descending intrusion, the flow will be steady in the co-moving frame. The steady version of the continuity equation,$ \grad \cdot(\rho{\bf u})=0$, is satisfied by a constant mass flux
\begin{equation}
\dot{m} = \rho u \alpha \label{continuity.eq}, 
\end{equation}
where $\rho$ is the density, $u$ is the speed of the fluid, $\alpha$ is the cross section of the flux tube defined by $f$ and $\dot{m}$ is a constant of integration that is conserved along ${\bf X}_f$. The steady flow must also satisfy the momentum equation
\begin{equation}
\rho ({\bf u} \cdot \grad ){\bf u}  = \frac{1}{4\pi}(\grad \times {\bf B}) \times {\bf B} + \grad \cdot \uuline{\sigma} - \grad p,
\end{equation}
where gravity is omitted for simplicity. Here $p$ is the plasma pressure and $\uuline \sigma$ is the viscous stress tensor. Since the flow must be parallel to the magnetic field, the Lorentz force makes no contribution to the parallel momentum equation
\begin{equation}\label{reduced_momentum.eq}
({\bf u} \cdot \grad) \frac{1}{2} u^2 + \frac{1}{\rho} ({\bf u} \cdot \grad ) p = \frac{1}{\rho}{\bf u} \cdot (\grad \cdot \uuline{\sigma}).
\end{equation}
This equation is the same as that of a neutral fluid passing though a nozzle. Together with an energy equation relating $\rho$ and $p$, Eqs.~\eqref{reduced_momentum.eq} and \eqref{continuity.eq} fully specify the spatial variation of the fluid along the length of the flux tube. 

For simplicity we adopt the isothermal equation of state 
\begin{equation}
p = C_s^2 \rho \label{isothermal.eq},
\end{equation}
where $C_s$ is the sound speed. This assumption is motivated by the very high thermal conduction along field lines. Combining Eq.~\eqref{isothermal.eq} with Eq.~\eqref{reduced_momentum.eq} and integrating over the volume of a fluid element with parameterized length $l$ leads to
\begin{equation}
\left . \left [ \frac{1}{2} u^2 + C_s^2 \ln \rho \right ] \right |_{l_1}^{l_2} = \int_{l_1}^{l_2} \mathrm{d}l ~\hat{u} \cdot \left [\frac{1}{\rho} \grad \cdot \uuline{\sigma} \right ], \label{energy.eq}
\end{equation}
where $l_1$ and $l_2$ are two arbitrary locations along the flux tube. Under strong magnetization the viscous force is dominated by the parallel contribution \citep{Guidoni_2010}. Using a field-aligned coordinate system it can be shown that this contribution takes the form
\begin{equation}\label{visc_stress.eq}
\hat{u}\cdot \left [ \grad \cdot \uuline{\sigma} \right ] = \frac{\mu^{(0)}}{\alpha^4} \left (\frac{4}{3}\partial_l^2  \left ( \frac{u}{\alpha} \right ) + 2 \partial_l \left (u ~\partial_l \frac{1}{\alpha} \right ) \right ),
\end{equation}
where $\mu^{(0)}$ is the dominant coefficient of dynamic viscosity, which is proportional to $\rho \lambda C_s$, and $\lambda$ is ion mean free path. When the flow is sufficiently smooth for the ion mean-free path to be negligible, the viscous contribution to the momentum equation may be neglected and the left hand side of Eq.~\eqref{energy.eq} is conserved along the length of the flux tube. Only in the case of a shock, where fluid variation cascades to shorter length scales, is the viscous contribution significant.

\begin{figure}[ht!]
\begin{center}
\includegraphics[width = 0.7\textwidth]{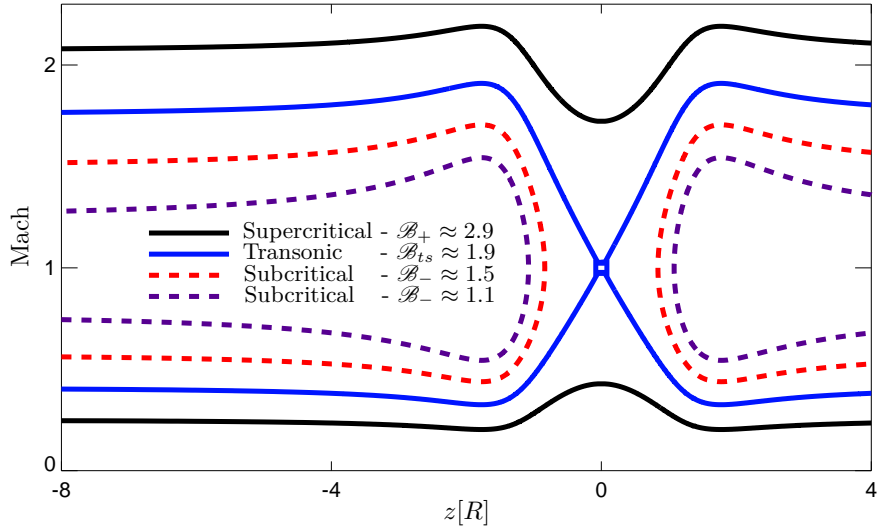}
\caption{\label{flow.fig} Contours of Eq.~\eqref{delaval.eq} are plotted for the $f = R/2$ field line. Each color represents a different value for $\mathscr{B}$. Supercritical solutions (black) are well-behaved. Subcritical solutions (purple, red) are ill-behaved where $M \rightarrow 1$ and ill-defined in the range $-z_{crit} < z < z_{crit}$. The two regimes are separated by the transonic contour (blue), which passes through $M = 1, z = 0$. Fluid flow is from left to right.}
\end{center}
\end{figure}

After neglecting viscocity, Eqs.~\eqref{energy.eq} and \eqref{continuity.eq} lead to a relationship between the flux-tube cross section $\alpha$ and the Mach number ($M = u / C_s$) that is equivalent de Laval's equation for steady flow through a nozzle;
\begin{equation}
M^2 -\ln M^2 -\ln \alpha^2 = \mathscr{B}.
\label{delaval.eq}
\end{equation} 
$\mathscr{B}$ is effectively Bernoulli's constant and is a conserved quantity along any time-independent, inviscid flow. $\mathscr{B}$ can, in principle, assume any real, positive value, and will generally be determined by evaluating $M$ and $\alpha$ at a particular point of interest. For real values of M the quantity $M^2 - \ln M^2$ has a minimum value of unity at $M = 1$ and diverges monotonically as $M$ goes to either zero or infinity. There are two solutions to Eq.~\eqref{delaval.eq} corresponding to any value of $\mathscr{B}$ -- one subsonic and one supersonic.

The behavior of this system can be visualized by plotting contours of $\mathscr{B}$ in $M$-$z$ phase space. Figure \ref{flow.fig} shows a representative set of solutions for the $f = R/2$ field line with $\mathscr{B}$ ranging from approximately 1.1 for the dashed purple contour up to nearly 3 for the black contour. The qualitative behavior of these solutions is dictated by how $\mathscr{B}$ relates to the critical value of 
\begin{equation}
\mathscr{B}_{ts}(f) = 1 - 2 \ln \alpha_{min}(f),
\end{equation}
which defines the transonic flow contour for which $M\rightarrow 1$ exactly at the throat of the constriction where $\alpha \rightarrow \alpha_{min}$. The transonic contour separates the so called {\it supercritical} solutions, given by $\mathscr{B} \in \mathscr{B_+} > \mathscr{B}_{ts}$, from the {\it subcritical} solutions, given by $\mathscr{B} \in \mathscr{B}_- < \mathscr{B}_{ts}$. As an example, consider the $f = R/2$ field line depicted in Figure \ref{flow.fig}. The minimum cross section is $\alpha_{min}(R/2) \approx 0.62$ and so $\mathscr{B}_{ts}(R/2) \approx 2$. Inverting Eq.~\eqref{delaval.eq} for $\alpha = 1$ (in the far field) we find the two transonic inflow values are $M_{ts}(R/2) \approx 1.75$ and $0.41$. 

Supercritical solutions have the property that $M_+(z)^2 - \ln M_+(z)^2 > 1$ for all values of $\alpha_f(z)$ so that $M(\mathscr{B}_+,f,z)$ is well defined along the entire flux tube. The black contours in Figure \ref{flow.fig} represent the supersonic and subsonic solutions for a particular value of $\mathscr{B}_+$. Note that these contours are everywhere either supersonic or subsonic and are well defined as $z\rightarrow 0$. Subcritical solutions do not have this property and are ill-defined at any location where the cross section is smaller than the so called {\it critical cross section}, given by 
\begin{equation}\label{alpha_crit.eq}
\alpha_{crit} = e^{(1 - \mathscr{B_-})/2}.
\end{equation}
Subcritical solutions are defined by the existence of a set of {\it critical points}, given by the two locations, $z = \pm z_{crit}(f)$, that satisfy Eq.~\eqref{alpha_crit.eq}. At these critical points $M(\mathscr{B}_-,f,z_{crit}) = 1$, while over the interval $-z_{crit} < z < z_{crit}$ the Mach number is ill-defined. The dashed red and purple contours of Figure \ref{flow.fig} represent subcritical solutions for two different values of $\mathscr{B}_-$.  In both cases the Mach number goes to unity at $z = \pm z_{crit}(f)$ and is ill-defined over the interval between the two critical points. A third solution is visible as the blue line in Figure \ref{flow.fig} and corresponds to the {\it transonic} contour with $\mathscr{B} = \mathscr{B}_{ts}$. This solution has the unique property that $\alpha_{crit} = \alpha_{min} (f)= \alpha_f(z = 0)$ so that the two critical points occur exactly at the throat of the flux tube. 

The two branches of the transonic contour separate the $M$-$z$ phase space into the subcritical region, which is located between the transonic contours, and the supercritical region, which is located above and below the supersonic and subsonic branches of the transonic solution, respectively. As with subcritical solutions, the supersonic and subsonic branches of the transonic solution eventually intersect as the Mach number goes to unity. But, unlike the subcritical contours, the transonic contour is well defined over the entire length of the flux tube. And since $\partial_z \alpha_f(z) = 0$ at $z = 0$, the fluid is well-behaved at this point even as the Mach number goes to unity. This contour is therefore the only viable solution that allows for a fluid to smoothly pass between supersonic and subsonic flows while conserving the value of $\mathscr{B}$.


\subsection{Transitional Flows}

If we were free to choose the value of $\mathscr{B}$ to be always equal to or greater than $\mathscr{B}_{ts}$ the steady solutions described in \ref{pflow.sec} would be sufficient. Since the fluid is at rest in the limb frame, in the intrusion frame the fluid velocity in the far field is given by ${\bf u}_{far} = u_{in} \hat{\bf z} = \hat{ \bf z} C_s M_{in}$, where $u_{in}$ is the speed at which the intrusion descends and $M_{in}$ is its Mach number. Since ($M_{in}$) must be allowed to assume any real value we are forced to consider that the far field boundary condition might correspond to a subcritical flow. The inadmissibility, between $-z_{crit}$ and $z_{crit}$, of a solution with the value of $\mathscr{B}$ fixed by the boundary condition demands that the overall solution be one in which $\mathscr{B}$ is not conserved.  This solution will take the form of several regions of constant $\mathscr{B}$, each connected by a transition in which Bernoulli's equation does not hold.  The transitions are either shocks or rarefaction waves whose locations change with time.  The complete flow combines two shocks, both propagating upstream, enclosing a transonic flow on which $\mathscr{B}=\mathscr{B}_{ts}$, and then a rarefaction wave propagating downstream away from the intrusion. For a careful discussion of shocks and rarefaction waves see Chapters IX and X of \cite{Landau_Lifshitz}. The following is a more specific discussion, aimed at our particular problem.


\subsubsection{Shocks}

In cases where the length scale of the fluid becomes comparable to the ion mean free path the viscosity has a non-negligible contribution to the momentum equation and cannot be ignored as it was leading to Eq.~\eqref{delaval.eq}. The resulting behavior is referred to as a shock, which is a thin transition from one value of $\mathscr{B}$ to another.  In a reference frame co-moving with the shock the flow must be steady and conserve mass and momentum.  In the isothermal case these conditions lead to a version of the Rankine-Hugoniot condition, amounting to conservation of
\begin{equation}
  M' ~+~\frac{1}{M'}
\end{equation}
across the jump, where $M'=(u-u_s)/C_s$ is the Mach number viewed from a frame moving at the shock speed $u_s$.  This conservation law differs from Eq.~\eqref{delaval.eq} because the jump is assumed so thin that viscosity cannot be ignored and $\alpha$ is approximately constant across it.  Discounting the trivial case where the Mach number is unchanged, leads to the
relation 
\begin{equation}\label{RH_isotherm.eq}
  M'_2 ~=~\frac{1}{M'_1}
\end{equation}
between upstream and downstream Mach numbers, $M'_1$ and $M'_2$.  It is evident that one of these will be subsonic while the other is supersonic.

In terms of Mach numbers $M_j$ in the frame of the intrusion, Eq.~\eqref{RH_isotherm.eq} takes the form
\begin{equation}
  M_2 ~-~M_s ~=~ \frac{1}{M_1-M_s} ~~,
\end{equation}
where $M_s=u_s/C_s$ is the Mach number of the shock.  Knowing the upstream and downstream Mach numbers then gives the 
shock Mach number as
\begin{equation}
  M_s ~=~ \frac{M_1+M_2}{2} - \sqrt{\left(\frac{M_1-M_2}{2}\right)^2 + 1},
\end{equation}
assuming $M_1>M_2>0$.  The shock will move leftward ($M_s<0$) if $M_1<1/M_2$, and rightward if $M_1>1/M_2$.  Mass conservation, in the shock reference frame, then leads to the relation
\begin{equation}
  \rho_2 ~=~ \rho_1\frac{M_1-M_s}{M_2-M_s} ~=~  \rho_1\frac{\sqrt{(M_1-M_2)^2+4}+(M_1-M_2)}{\sqrt{(M_1-M_2)^2+4}-(M_1-M_2)}~~,
\end{equation}
between pre-shock and post-shock density.  A shock must have $M_1^\prime>1>M_2^\prime$, and therefore $\rho_2>\rho_1$: it is compressive.


\subsubsection{Rarefaction waves}

A jump to lower density, not possible in a shock, must occur in a rarefaction wave.  In cases without externally defined length scale the rarefaction wave will be {\it self-similar} \citep[\S92]{Landau_Lifshitz}, depending on space and time through a single similarity variable $(z-z_0)/t$. A rarefaction wave is inherently time-dependent and so Bernoulli's equation is again invalid. In our solution, a shock and rarefaction wave will be generated simultaneously at $t=0$ from the single point $z=z_{crit}$.  This initial state lacks a length scale and we may take the downstream rarefaction wave to be of the self-similar form.  It will be bounded by weak discontinuities at its edges.  The leading edge, at $z=z_2$, propagates into the (unperturbed) downstream plasma at $u_2+C_s$.  Upstream of this the velocity, and thus Mach number, is linear \citep{Landau_Lifshitz}
\begin{equation}
  M ~=~ M_2 - \frac{z_2-z}{C_s t}.
\end{equation}
Upstream of the trailing edge, at $z=z_1$, the flow is again constant with $M=M_1<M_2$.  Thus the extent of the rarefaction wave grows in time as $\Delta z=(M_2-M_1)C_s t$, beginning as a discontinuity at $t=0$.  The initial discontinuity at $z=+z_{crit}$ decomposes into this rarefaction wave and a shock, in the manner of a Riemann problem \citep{Landau_Lifshitz}.

Within the rarefaction wave the density is an explicit function of velocity \citep[see][\S 92]{Landau_Lifshitz}
\begin{equation}
\rho ~=~ \rho_2\,e^{M-M_2}~=~ \rho_2\,\exp\left(\frac{z-z_2}{C_s t}\right).
\end{equation}
The upstream and downstream densities are therefore
related by
\begin{equation}
\rho_2 ~=~\rho_1\,e^{M_2-M_1},
\end{equation}
across the ever-expanding rarefaction wave. In order for this solution to apply the interior size of the rarefaction wave must be much smaller than the length scale of variation of the fluid cross-section, $\alpha$. Fortunately, the rarefaction wave, while growing in time, propagates vary quickly into the far field so that no matter how large it becomes, the scale over which $\alpha$ varies is always larger still.


\subsection{Composite Flow} \label{comp_flow.sec}

The complete solution, defined over the entire length of the affected flux tube, will be piecewise continuous using shocks, rarefaction waves and regions of pertistaltic flow so that the fluid velocity and density are treated in an internally consistent manner. The locations of the transition flows will travel along the length of the flux tubes in order to satisfy their governing equations and will therefore introduce time variations into the system despite our previous assumption of time independence. In letting this system evolve we are assuming that it can be treated as an adiabatic series of time-independent solutions. This assumption will be valid so long as the timescale over which a given feature evolves is long compared to its fluid crossing time.  

In order to form a solution we use physical consideration to motivate the choice of initial conditions in the region between the critical points. Far above and below the intrusion we demand that the plasma density and velocity be unchanged and continuity demands that every jump in velocity have a corresponding jump in density. We therefore require at least two jumps with at least one unspecified intermediate value of $\mathscr{B}$ in order to have sufficient degrees of freedom to satisfy the boundary conditions on both velocity and pressure, which is equivalent to density in the isothermal limit. 

If the fluid velocity in the far field is supersonic and lies below the transonic contour, then $\mathscr{B}_{in}$ is subcritical and $M$ will be ill-defined between the two critical points.\footnote{$M_{in}$ could also be subsonic and above the lower transonic contour. For now we consider only supersonic inflows.} In order for the fluid to avoid an infinite acceleration at the upstream critical point there must be a transition away from $\mathscr{B}_{in}$ and onto some well-behaved flow, $\mathscr{B} \in \mathscr{B}_+$. In order for solution to remain well defined the transition must propagate upstream, away from the critical point. This is only possible in the case of a shock. A rarefaction wave would not propagate upstream with sufficient speed to escape the critical region. The downstream flow must therefore be subsonic in order for the system to be well defined at the upstream critical point.

At the downstream critical point there must again be a transition to connect the flow that resulted from the upstream shock back to the original contour $\mathscr{B}_{in}$. If the interior flow were everywhere subsonic the jump would again have to be a rarefaction wave and would propagate at Mach 1 into the higher density, subsonic fluid, ultimately making its way into the critical region and leaving the downstream critical point again ill-behaved. Thus, in the downstream region, the flow that resulted from the upstream shock must be supersonic at the critical point. Only the transonic contour $\mathscr{B}_{ts}$ can satisfy this condition without introducing yet another jump within the critical region. We therefore reach the conclusion, well known in nozzle problems, that the flow must cross from subsonic to supersonic at the throat.

\begin{figure}[ht!]
\begin{center}
\includegraphics[width = 0.7\textwidth]{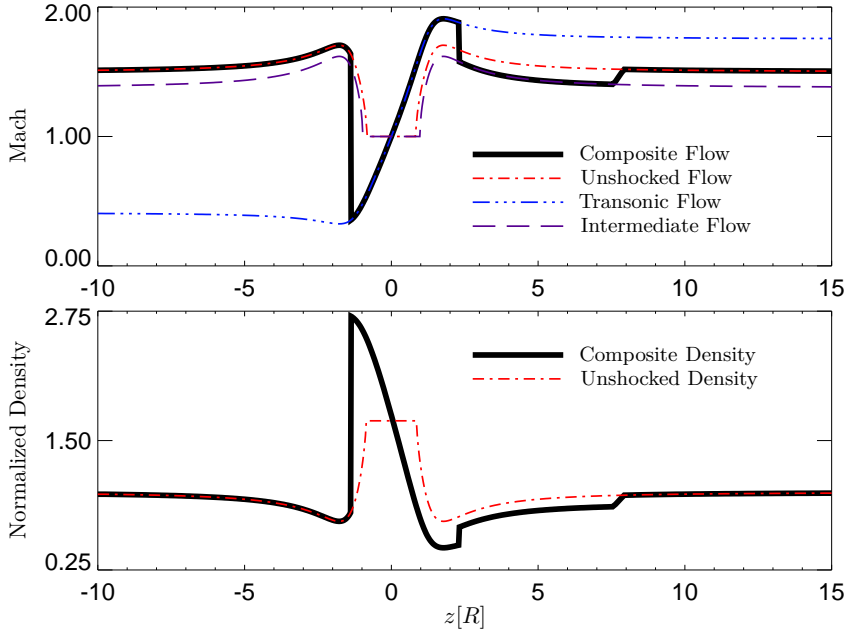}
\caption{\label{ps_flow.fig} The piecewise continuous, composite flow is formed by connecting the subcritical flow from the far field with the transonic solution in the interior and an unspecified subcritical solution in the intermediate downstream region. The unshocked density and Mach number indicate the far field subcritical solution in the absence of shocks. As in Figure \ref{flow.fig} fluid flows from left to right.}
\end{center}
\end{figure}

The density change across the leading shock, which connects the subcritical inflow to the transonic interior flow, is fixed by the relative speeds of the fluid on either side of the shock. At the downstream critical point the speed of the fluid is given by the transonic solution and the density is fixed through the continuity equation. Any transition from the transonic flow back to the original flow must therefore satisfy the disparity in both speed and density at this point, a feat not achievable for either a shock or a rarefaction wave. The jump at the downstream critical point must therefore decompose into both a shock and a rarefaction wave, just as in an asymmetric Riemann problem \citep{Landau_Lifshitz}. The rarefaction wave then propagates at Mach 1 into the downstream fluid and therefore moves away from the intrusion at speeds in excess of Mach 2. The shock propagates subsonically upstream into the transonic flow, which is itself supersonic, and therefore moves more slowly away from the intrusion. 

Between the downstream shock and the rarefaction wave there is an initially infinitesimal intermediate region in which the fluid lies on an unspecified contour of $\mathscr{B}$, which will be determined such that the net effect of the two downstream transitions exactly compensates for the upstream shock and transonic interior. The intermediate flow is supersonic but also slower than the initial, subcritical flow, so it too has critical points and these must be accounted for when the initial locations of the three transitions are chosen. The whole system is evolved by using the current velocity of each feature to determine its location at some future time and then constructing the new velocity and density profiles for the whole system at that time. This construct is shown in Figure \ref{ps_flow.fig} for the $f = 0.5$ field line with an inflow condition of $M_{in} = 1.5$. The system is shown a short time after launch so that the transitions are spatially separated and can be easily distinguished. The original, subcritical flow is unphysical at $z \approx -R$ but the composite flow is transonic at this point and thus exhibits no critical phenomena.

Relative to the intrusion, the upstream transonic flow and downstream intermediate flow are both slower than the subcritical flow in the far field. In the limb frame the plasma in these regions is actually descending toward the limb along with the intrusion. The two shocks similarly descend toward the limb with the leading shock pushing ahead and the trailing shock lagging ever farther behind while the rarefaction wave is sufficiently fast that it is not entrained with the intrusion and escapes rapidly upward. Note however that these shocks are not standoff shocks. They evolve in time and move steadily away from the intrusion, ultimately finding their way into the far field where their evolution slows and the assumption of time-independence becomes increasingly exact.


\section{Results}

\begin{figure}[hb!]
\begin{center}
\includegraphics[width = 0.7\textwidth]{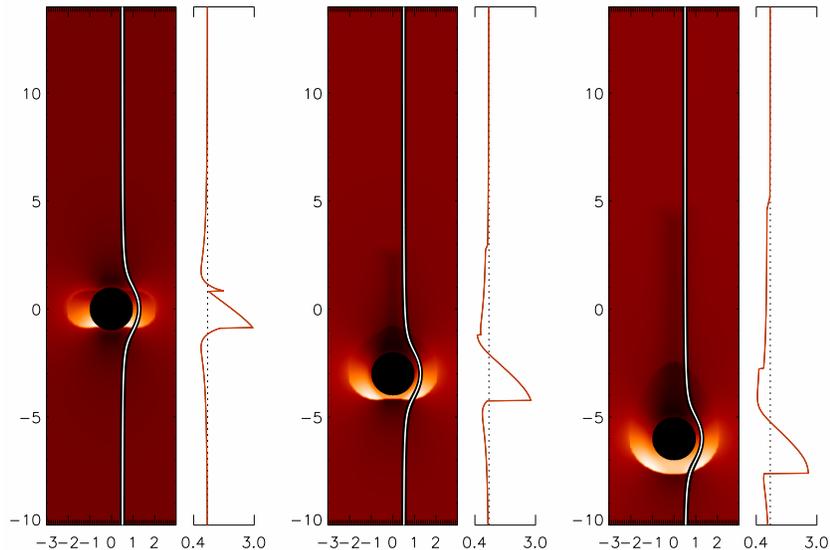}
\caption{\label{tri_2d.fig} A $M_{in} = 1.5$ descending intrusion is shown in the limb frame at times $t = \{0,3,6\} R/C_s$. The high and low density regions are seen in red scale and the plots to the right of each panel show the exact density profile for the field line traced in white. The dashed lines indicate a normalized density of 1, as in the far field.}
\end{center}
\end{figure}

In order to gain insight into solar dynamics from this model we must construct synthetic observables which can be compared to actual observations. To this end we begin by constructing 2D maps of density and velocity, made by interpolating between a representative sample of field lines, each determined with the same boundary conditions. Features of the 1D fluid solution manifest in the 2D maps as broad fronts, and regions of high or low density. Figure \ref{tri_2d.fig} shows one such map of plasma density given at three successive times. From this example several features are visible. The leading shock, trailing shock and rarefaction wave are all distinguishable as abrupt changes in the plasma density. The high and low density ``head'' and ``tail'' grow in time and descend toward the limb while a slightly less rarefied region between the trailing shock and the rarefaction wave grows quickly upward as the rarefaction wave escapes away from the limb.

\subsection{Emission Measure}

\begin{figure}[ht]
\begin{center}
\includegraphics[width = 0.7\textwidth]{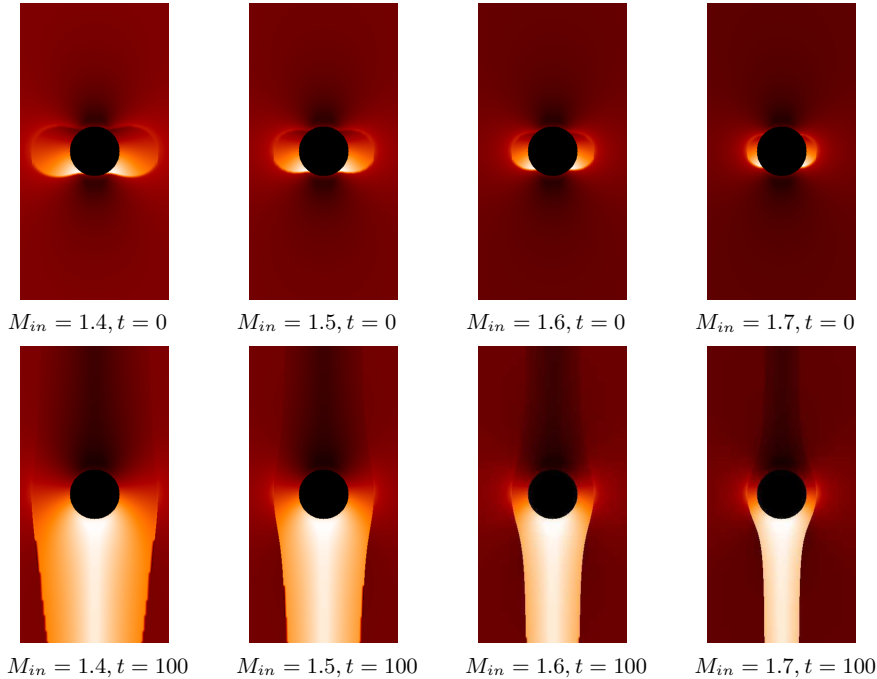}
\caption{\label{comp_quad.fig} Four peristaltic shocks are shown for $M_{in} = \{1.4, 1.5, 1.6, 1.7\}$ (left to right) and for times $t = \{0, 100\}R/C_s$ (top to bottom). The edges of the shocked column in the lower row trace out the field lines that are transonic for each value of $M_{in}$, which separate the shocked and unshocked regions. }
\end{center}
\end{figure}

As a proxy for synthetic images of the optically thin corona we choose the emission measure density ($\epsilon \propto \rho ^2$). The emission measure profile will depend on the viewing angle. To begin with we consider a line of sight that is normal to the current sheet, consistent with many imaging observations of sheet-like structures above post-CME solar arcades \citep{Svestka_1998, Gallagher_2002, Innes_2003, Savage_2011, Savage_2012, McKenzie_2013}. If the intrusion pierces normally through the current sheet as in Figure \ref{cartoon1.sub} then $\epsilon$ can be constructed simply by squaring the 2D density maps such as in Figure \ref{tri_2d.fig}. This viewing angle also applies to cases where the intrustion is imbedded within the current sheet (as in Figure \ref{cartoon2.sub}), which is itself viewed edge on. The resulting emission measure maps will exhibit the same features as Figure \ref{tri_2d.fig}. A collection of four such emission measure maps is shown in Figure \ref{comp_quad.fig} for four different descending intrusions, each depicted at the instant of launch and then again after the shocks have propagated into the far field. At $t = 0$ the shocks trace out the loci of critical points for each field line. Then, as $t \rightarrow \infty$ the shock fronts move into the far field so that the shock column has infinite vertical extent both above and below the intrusion. 

Relative to the diameter of the intrusion, the width of the shocked column depends only on the speed of the intrusion $M_{in}$, which dictates the fluid velocity in the far field. For intermediate speeds (between Mach 1.4 and and 1.7) the column width is of the same order as the intrusion diameter. The upper limit occurs as $M_{in} \rightarrow 1.92$ at which point all field lines become non-critical so the shocked column vanishes. In the opposite limit, as $M_{in} \rightarrow 1$ all field lines exhibit critical behavior and the shocked column becomes infinite in width but with vanishing amplitude in the far field. 

\begin{figure}[ht!]
\begin{center}
\includegraphics[width = 0.45\textwidth]{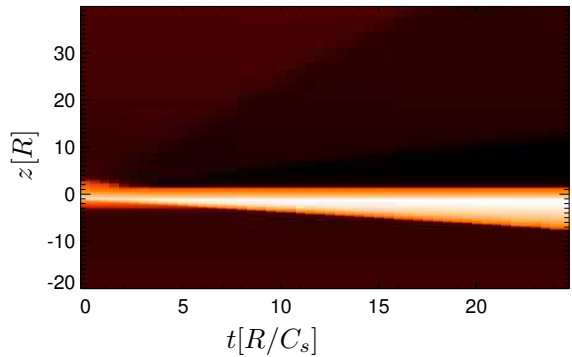}
\caption{\label{LOS.fig} The line-of-sight integrated emission measure depicted here as a stack plot. Higher emission is indicated in yellow. Time increases to the right with the vertical profile at any given time corresponding to $\epsilon(z,t).$}
\end{center}
\end{figure}

As an alternative we consider the system viewed from the side, such as if the line of sight were along $\hat{\bf y}$ in Figure \ref{cartoon2.sub}. In this case the emission measure is constructed by integrating transversely across the 2D domain. The resulting profile will resemble that of an individual field line but will be somewhat smoother, having effectively averaged over all shocked field lines. We define the background-subtracted, normalized column emission measure as
\begin{equation}
\epsilon(z) = \int_{-L}^{L} \mathrm{d} y \left ( \frac{\rho(y, z)^2}{\rho_0^2}  -1\right ),
\end{equation} 
so that $\epsilon(z)$ goes to zero for $\rho(y,z) = \rho_0$ and is negative or positive where $\rho$ is depleted or enhanced with respect to the ambient plasma. Figure \ref{LOS.fig} shows a stack-plot of successive time-steps of $\epsilon(z,t)$ for $M_{in} = 1.3$. The propagating shocks and rarefaction wave can be seen as abrupt jumps in the red scale emission.

\begin{figure}[ht!]
\begin{center}
\includegraphics[width = 0.45\textwidth]{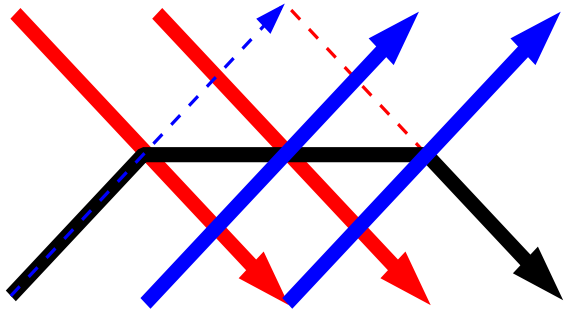}
\caption{\label{shear.fig} A sheared magnetic field results in a reconnected, horizontal field line which is drug downward through the adjacent layers of unreconnected field, forming a plateau.}
\end{center}
\end{figure}

To visualize how this kind of structure might manifest in the current sheet consider the case of a slightly sheared supra-arcade magnetic field. According to \cite{Guidoni_2010}, a local reconnection event will result in a growing, descending trapezoidal plateau that leads to something like Figures \ref{cartoon2.sub} and \ref{shear.fig}. The horizontal segment of the reconnected field, i.e. the intrusion, is embedded in the current sheet and drives peristaltic flows in the surrounding layers of field. As the plateau descends the bends move outward so that more of the reconnected flux is embedded in the current sheet.

Because the field is sheared, we treat the two respective layers of magnetic field independently. They both exhibit peristaltic flows which result in an emission such as in Figure \ref{LOS.fig}, but in one field the flow is slanted slightly to the right while the other is slanted to the left. The composite flows launch first on the field lines closest to the initial reconnection point. Then, as the plateau descends, those same field lines continue to evolve while field lines that are newly exposed to the growing plateau are initiated each in turn. The net result is a locus of shocked flows that grows as the plateau grows. 

\begin{figure}[ht!]
\begin{center}
\includegraphics[width = 0.45\textwidth]{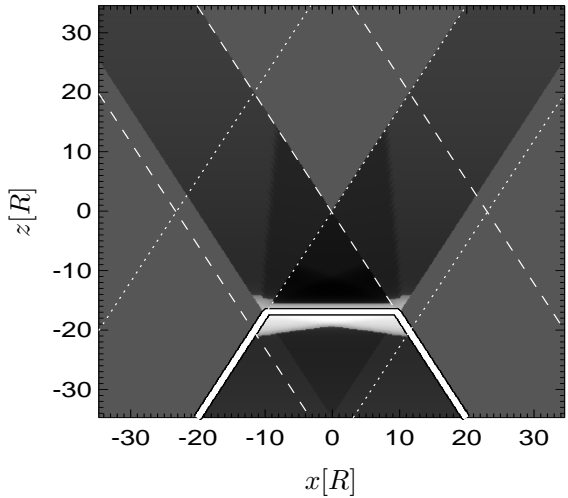}
\caption{\label{wedge.fig} A high contrast plot of emission measure for a sheared peristaltic event with two sets of field lines, dashed and dotted. The locus of shock features are visible above and below the plateau created by the embedded segment of reconnected flux, depicted as a solid white line.}
\end{center}
\end{figure}

Such a system is depicted in Figure \ref{wedge.fig}. The unreconnected field is angled up and to the right in the foreground and down and to the right in the background. The layers of field that pass close to the horizontal segment of reconnected field exhibit peristaltic pumping. The emission measure on each field line is given by $\epsilon(\tilde{z}, t - t_f)$, where $\tilde{z}$ is the distance along the angled field line and $t_f$ is the time at which that field line was initiated. The collection of lower shocks leads to a vaguely arch shaped high density region while the upper shock leads to a similar rarefied region. The antisunward rarefaction waves lead to a nearly vertical column of low density owing to the fact that these waves propagate supersonically outward along field lines at a rate comparable to the growth of the plateau.

\subsection{Momentum in the Fluid}

\begin{figure}[ht!]
\begin{center}
\includegraphics[width = 0.5\textwidth, trim = 10 0 -10 0, clip = true]{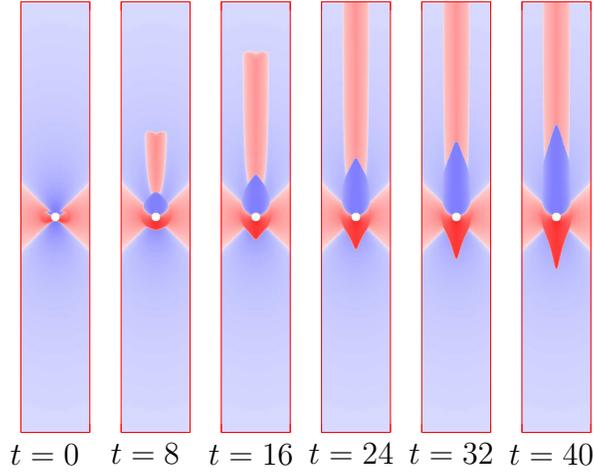}
\caption{\label{mom_dens.fig} A time series of fluid momentum density. Red indicates sunward momentum while blue is antisunward.}
\end{center}
\end{figure}

Since the descending intrusion generates plasma motion in the surrounding field, there should be associated energy and momentum transfer into that fluid. For any finite domain we can find the momentum in the plasma through numerical integration of the plasma density and velocity. Figure \ref{mom_dens.fig} displays a representative plot of momentum density. 

Due to the fact that the normalized cross section in the far field is asymptotic to but always greater than unity, the supersonic fluid far from the intrusion always propagates slightly faster than $M_{in}$ in the rest frame of the intrusion. In the limb frame this fluid is slowly rising so the momentum density far from the intrusion is always anti-sunward. Closer to the intrusion it is directed sunward as the cross section becomes constricted and the fluid is slowed below $M_{in}$. Immediately above the lower shock the momentum density is strongly sunward and then becomes anti-sunward as the transonic flow passes abreast of the intrusion and again exceeds $M_{in}$. It then becomes sunward again across the second shock before finally returning to the far-field limit across the rarefaction wave.

\begin{figure}[ht!]
\begin{center}
\includegraphics[width = 0.45\textwidth]{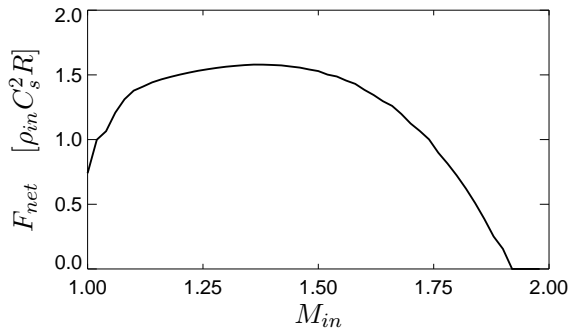}
\caption{\label{o_fluid_force.fig} The time rate of change of momentum in the fluid yields a net force on the fluid that vanishes when the width of the shocked column goes to zero as $M_{in} \rightarrow 1.92$ and remains finite as the width expands and $M_{in} \rightarrow 1$.}
\end{center}
\end{figure}

To explore this more carefully we observe that the force per unit fluid cross section on a shocked flux tube may be found explicitly through 
\begin{align}
f = & \partial_t \int_{-\infty}^{\infty} \mathrm{d}{\bf z} \cdot (\rho {\bf u} \alpha) \\
  = & \partial_t \left(\int_{z_1-\delta}^{z_1+\delta} + \int_{z_2 - \delta}^{z_2+\delta} + \int_{z_3 - \delta}^{z_3+\delta} \right)\mathrm{d}{\bf z} \cdot (\rho {\bf u} \alpha)\\
  = & v_{z_1} \left . \left [ \rho u \alpha \right ] \right |_{z_1}^{z_1 + \delta_1} + v_{z_2} \left . \left[ \rho u \alpha \right ] \right |_{z_2}^{z_2 + \delta_2} + v_{z_3} \left . \left [\rho u \alpha \right ] \right |_{z_3}^{z_3 + \delta_3}~,
\end{align}
where $\delta_i$ represents the width of each jump. This can be calculated numerically for every field line within a finite domain under the assumption that all three jumps have propagated into the far field where the shock speeds become steady and $\alpha \rightarrow 1$. In Figure \ref{o_fluid_force.fig} we see that the net force on the fluid appears to be finite even in the limit of $M\rightarrow 1$, where the shock becomes infinitely wide. In order to confirm this the contribution from the far field may be approximated analytically. This calculation is not included in the present work but it can be shown that the force contributed by the shocks in the far field vanishes with an inverse power of distance greater than unity so that, indeed, the net force on the fluid remains finite even as the shocks fill all of space. 

\subsection{Drag Force}

If the total momentum in our model contained only two contributing terms we could use Figure \ref{o_fluid_force.fig} as a proxy for the drag force on the intrusion. In actuality the far-field boundary conditions also contribute momentum to the system and the drag force must be calculated explicitly by integrating the plasma pressure over the surface of the intrusion. The pressure is related to the density, which can be found explicitly by calculating the behavior of fluid on the $f = 0$ streamline. The net vertical force due to the pressure $p_s(\theta)$ on the surface $S$ is then
\begin{equation}
F_z = 2 \int_0^\pi r \mathrm{d} \theta ~ p_s \hat{\bf n} \cdot \hat{\bf z} = 2 \int_0^\pi p_s(\theta) r \cos(\theta) \mathrm{d} \theta~.
\end{equation}

\begin{figure}[hb!]
\begin{center}
\includegraphics[width = 0.45\textwidth]{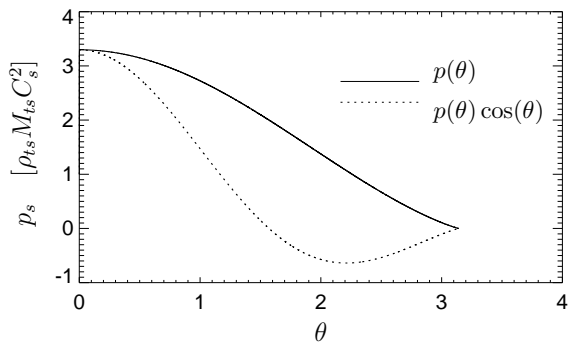}
\caption{\label{pressure.fig} Pressure along the surface of the intrusion, parameterized by polar angle. $\rho_{ts}$ and $M_{ts}$ are the transonic density and Mach number far below the intrusion. The integrated pressure gives the force on the intrusion (per unit inserted length) as $ F \approx 4.8 \times \rho_0 M_0 R C_s^2$.}
\end{center}
\end{figure}

The plasma on the $f = 0$ field line is always on the transonic contour. The $f = 0$ streamline intersects the surface of the intrusion at the two magnetic null points $\theta = \{ 0, \pi \}$, with $\theta$ measured here from $-\hat{\bf z}$. At these points the fluid cross section diverges as the magnetic field strength vanishes. For $ \theta = \pi $ the fluid Mach number also diverges so $\rho$ must be zero by continuity. For $\theta = 0$ the plasma is subsonic so $M$ goes to zero as $\alpha \rightarrow \infty$. To ensure that the density is well-behaved at this point we perform a series expansion around $\theta = 0$ and find that $\rho (\theta \rightarrow 0) \approx 3.3 \rho_{ts}$, where $\rho_{ts}$ is the plasma density on the transonic contour far from the intrusion. $p(\theta)$ is therefore well-behaved and can be calculated numerically as in Figure \ref{pressure.fig}. The pressure decreases monotonically indicating an upward net force on the intrusion. The $z$-component of this force is given explicitly by the area under the dashed curve.

\begin{figure}[hb!]
\begin{center}
\includegraphics[width = 0.45\textwidth]{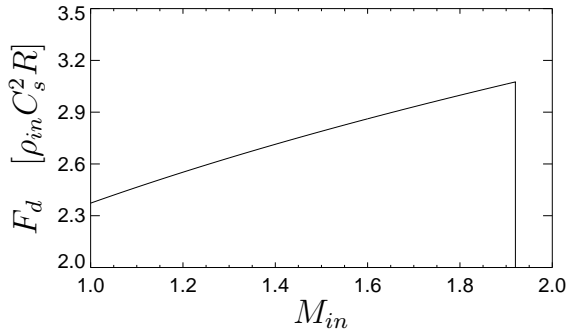}
\caption{\label{drag.fig} The drag force is found by evaluating the density jump across the lower shock for a given intrusion Mach number and then combining the result with the integration factor from Figure \ref{pressure.fig}.}
\end{center}
\end{figure}

In order to find how the drag force depends on $M_{in}$ we must find the pressure jump across the leading shock, which will dictate $\rho_{ts}$ and hence $p_{ts}$. The shock velocity $M_s$ is determined by $M_{in}$ and $M_{ts}$. But, in the far field $M_{ts} \rightarrow 0.319$ for $f = 0$, $\alpha = 1$. Thus, $\rho_{ts}$ depends only on $M_{in}$ and, when multiplied by the integration factor from Figure \ref{pressure.fig}, the resulting drag force can be found as depicted in Figure \ref{drag.fig}. The drag is lowest for the Mach 1 limit and increases almost linearly up to $M_{in} \approx 1.92$, at which point the column disappears. At and above Mach 1.92 the drag is zero due to the symmetry of the de Laval flow solutions, just as in D'Alembert's paradox. Below Mach 1 we have not calculated the drag profile but we expect, given the extent of the subsonic critical regime, that the drag will remain finite down to the minimum critical value of $M_{min} \approx 0.32$, at which point the shocked column again vanishes. 

As indicated, the drag curve in Figure \ref{drag.fig} does not match the net force on the fluid from Figure \ref{o_fluid_force.fig}. This is not surprising since the transonic flow is assymetric along the vertical direction and thus the centripetal force on each fluid element is unbalanced. Thus the magnetic field must deform asymmetrically in order to balance the fluid pressure. It follows that there must be some infinitesimal asymmetry between the far field magnetic field above and below the intrusion which in turn leads to a net force exerted on the system by the far field boundary conditions.


\section{Discussion}

In this work we have shown how field line retraction following a local reconnection event can manifest as a descending constriction in the nearby unreconnected field. This constriction behaves in many respects as a peristaltic pump, which leads to peristaltic flows and ultimately to shocks and rarefaction waves that alter the velocity and density of plasma on affected field lines. These shocks are not to be confused with standoff shocks, which form at a fixed distance in front of a traveling obstacle and are thereafter stationary in time. The fluid jumps that we have described cannot exist as time-independent solutions and must necessarily propagate away from the intrusion. The region between these jumps therefore grows in time and is, in and of itself, a dynamic feature.

The composite flows that form in this model are restricted to a column whose width is defined by the field-line that exhibits transonic flow for a particular boundary condition, $M_{in}$. This width increases monotonically as the speed of the intrusion decreases toward the sound speed. The minimum width of zero occurs when the speed of the descending intrusion reaches $M_{max} = 1.92$ while the maximum width is arbitrarily large as $M_{in} \rightarrow 1$. In this analysis we considered only supersonic values for $M_{in}$ but we acknowledge that the peristaltic flows will continue to exhibit critical behavior even for subsonic values of $M_{in}$ down to the point where $\mathscr{B}$ is again larger than $\mathscr{B}_{ts}$ on the $f = 0$ field line. 

Our profiles for the 2D density and emission measure maps bear striking resemblance to observations of voids and Supra-Arcade Downflows (SADs) in post-CME flares \citep{McKenzie_2000, Savage_2012}. In our analysis we considered an isothermal plasma in order to make the development more tractable. We offer, without proof, that an adiabatic plasma would exhibit the same qualitative behavior with the addition that plasma in the region between the lower shock and the rarefaction wave would exhibit an increase in temperature. The rarefied tail and high-emission leading edge may even be useful as thermal diagnostics since a temperature increase in the rarefied column could move the emission outside of a particular observation band-pass, thereby increasing the contrast in these features. 

We also considered an alternate geometry in which the particular shape of the emission profile is exchanged for a column integrated emission measure which occurs everywhere along the length of an embedded flux tube. This geometry also offers an interpretation for down-flowing features but may be more accurately used to describe how reconnection events and the contraction of reconnected flux can lead to heating of plasma along a broadly distributed volume of unreconnected field. 

Our model also helps to explain how a thermal halo \citep{Seaton_2009} might form around the current sheet. Here we have described only the creation of shocks along constricted field lines. But these shocks could very well travel down the unreconnected field all the way to the chromosphere where they would then drive evaporation exactly as in conduction dominated flare loops \citep{Cargill_1995}. This evaporation might then increase the density on ``post-peristaltic'' field lines, which could then undergo their own reconnection or even experience another ``peristaltic process'' in the event of a second nearby reconnection event. 

We further describe how the alterations to velocity and density relate to the momentum density in the fluid and the subsequent net force (per unit embedded length) on the fluid. This force is related to but not equal to the net force on the retracting flux tube since a third contribution comes from the boundary conditions which maintain the field profile in the far field. The net force on the intrusion is found from a direct calculation of pressure integrated over its surface. This force points in the direction opposite the motion of the intrusion and is of order $R C_s^2 \rho_{in}$. It increases almost linearly by nearly a factor of two over the range $1 < M_{in} < 2$. Larger descent speeds correspond to a larger drag force so that, if this force is sufficient to influence the kinematics of the descending intrusion, the drag will decrease as the intrusion slows. 

This drag force offers a possible explanation to the fact that reconnection outflows appear to move sub-Alfv\'enically despite the predictions of reconnection models such as described by \cite{Seaton_2009}. If reconnection outflows originate in locations where peristaltic shocks can form then this could lead to a drag force that would keep their velocities below the Alfv\'en speed. However, if the outflow velocity ever exceeds the maximum shock velocity then the drag force should vanish according to our model. These loops would descend much more rapidly in its absence. This line of reasoning suggests there may be a bimodal distribution in the velocity of retracting magnetic loops. Loops that move fast enough to avoid launching peristaltic shocks would remain fast-moving while slower moving loops would be damped by the momentum transferred into the plasma. 

In order for the aforementioned drag force to have a non-negligible influence, the plasma pressure must be comparable to the magnetic energy density in the retracting flux tube. But pressure balance between the intrusion and its surroundings demands that the field strength must be comparable between the retracting flux and the unreconnected field. Thus, the drag force will only be significant if the plasma pressure is comparable to the magnetic energy density in the unreconnected field. While this is not unlikely in reality it is in conflict with the zero-$\beta$ assumption and therefore cannot be reconciled with our model in its current form. 

Our model assumes an extremely low $\beta$ value in order to invoke the rigid magnetic field. However, observations suggest that this may not be an accurate assumption in the supra-arcade region \citep{McKenzie_2013}. It may be that by the time supra-arcade downflows become visibile in observations the local plasma $\beta$ has already been increased due to previous instances of peristaltic pumping and that our model only applies to the early stages of flare activity, when the plasma density and temperature are still relatively low. Future work will therefore require the relaxation of the low $\beta$ approximation, which will necessitate a numerical simulation. Another issue with the model is that we have been forced to stitch together time-independent solutions in an adiabatic fashion. The validity of this approximation, as well as those used in deriving the 1D MHD simplifications, will likewise need to be tested through simulations. 

When comparing to observations, some key differences are also apparent. Our model cannot reproduce the oscillatory behavior on the edges of voids as seen in \cite{Verwichte_2005}, although it does predict a discontinuity in plasma density transverse to the field, which could support surface modes if the zero-$\beta$ assumption were relaxed. Also, while we predict that these features should occur for $M_{in} \lessapprox 2$, \cite{Savage_2011} measured a typical downflow speed of on the order of $10^2$ km s$^{-1}$ with some instances of much higher values. Depending on the local sound speed these velocities may fall above our Mach 2 prediction. A more careful study of SAD speeds and the associated local sound speed will need to be conducted in order to refine this estimate. Moreover, the upper limit yielded by our simple model can be relaxed by generalizations to non-circular intrusions. As a first attempt we calculated that for elliptical intrusions the value of $M_{max}$ could be increased by a factor of nearly two before the aspect ratio of the ellipse became unrealistic. It may be that an appropriate choice of intrusion cross section could reconcile any lingering disparities between the model and observations of flare loops. Ultimately we intend to further the investigation with a regimen of numerical simulations. The ultimate success of this model will be in providing a theoretical framework for interpreting features seen in more complex numerical simulations.

\section{Acknowledgements}

The authors are grateful to K. Reeves and S. Savage for valuable comments on early drafts of the manuscript.  This work was supported in part by NASA under contract NNM07AB07C with the Smithsonian Astrophysical Observatory and in part by a grant form the NSF/DOE Partnership in Basic Plasma Physics.


\begin{thebibliography}{}
\expandafter\ifx\csname natexlab\endcsname\relax\def\natexlab#1{#1}\fi

\bibitem[{{Cargill} {et~al.}(1996){Cargill}, {Chen}, {Spicer}, \&
  {Zalesak}}]{Cargill_1996}
{Cargill}, P.~J., {Chen}, J., {Spicer}, D.~S., \& {Zalesak}, S.~T. 1996, \jgr,
  101, 4855

\bibitem[{{Cargill} {et~al.}(1995){Cargill}, {Mariska}, \&
  {Antiochos}}]{Cargill_1995}
{Cargill}, P.~J., {Mariska}, J.~T., \& {Antiochos}, S.~K. 1995, \apj, 439, 1034

\bibitem[{{Cassak} {et~al.}(2013){Cassak}, {Drake}, {Gosling}, {Phan}, {Shay},
  \& {Shepherd}}]{Cassak_2013}
{Cassak}, P.~A., {Drake}, J.~F., {Gosling}, J.~T., {et~al.} 2013, ArXiv
  e-prints, arXiv:1307.3946

\bibitem[{Choudhuri(1998)}]{Choudhuri}
Choudhuri, A.~R. 1998, The Physics of Fluids and Plasmas, An Introduction for
  Astrophysicists (New York: Cambridge University Press)

\bibitem[{{Ciaravella} \& {Raymond}(2008)}]{Ciaravella_2008}
{Ciaravella}, A., \& {Raymond}, J.~C. 2008, \apj, 686, 1372

\bibitem[{{Cliver} \& {Hudson}(2002)}]{Cliver_2002}
{Cliver}, E.~W., \& {Hudson}, H.~S. 2002, Journal of Atmospheric and
  Solar-Terrestrial Physics, 64, 231

\bibitem[{{Costa} {et~al.}(2009){Costa}, {Elaskar}, {Fern{\'a}ndez}, \&
  {Mart{\'{\i}}nez}}]{Costa_2009}
{Costa}, A., {Elaskar}, S., {Fern{\'a}ndez}, C.~A., \& {Mart{\'{\i}}nez}, G.
  2009, \mnras, 400, L85

\bibitem[{{Gallagher} {et~al.}(2002){Gallagher}, {Dennis}, {Krucker},
  {Schwartz}, \& {Tolbert}}]{Gallagher_2002}
{Gallagher}, P.~T., {Dennis}, B.~R., {Krucker}, S., {Schwartz}, R.~A., \&
  {Tolbert}, A.~K. 2002, \solphys, 210, 341

\bibitem[{{Guidoni} \& {Longcope}(2010)}]{Guidoni_2010}
{Guidoni}, S.~E., \& {Longcope}, D.~W. 2010, \apj, 718, 1476

\bibitem[{{Innes} {et~al.}(2003){Innes}, {McKenzie}, \& {Wang}}]{Innes_2003}
{Innes}, D.~E., {McKenzie}, D.~E., \& {Wang}, T. 2003, \solphys, 217, 247

\bibitem[{{Ko} {et~al.}(2010){Ko}, {Raymond}, {Vr{\v s}nak}, \&
  {Vuji{\'c}}}]{Ko_2010}
{Ko}, Y.-K., {Raymond}, J.~C., {Vr{\v s}nak}, B., \& {Vuji{\'c}}, E. 2010,
  \apj, 722, 625

\bibitem[{{Landau} \& {Lifshitz}(1959)}]{Landau_Lifshitz}
{Landau}, L.~D., \& {Lifshitz}, E.~M. 1959, Fluid Mechanics, Landau and
  Lifshitz Course of Theoretical Physics, Volume 6 (New York: Pergamon Press
  Ltd)

\bibitem[{{Linton} \& {Longcope}(2006)}]{Linton_2006}
{Linton}, M.~G., \& {Longcope}, D.~W. 2006, \apj, 642, 1177

\bibitem[{{Longcope} {et~al.}(2009){Longcope}, {Guidoni}, \&
  {Linton}}]{Longcope_2009}
{Longcope}, D.~W., {Guidoni}, S.~E., \& {Linton}, M.~G. 2009, \apjl, 690, L18

\bibitem[{{McKenzie}(2000)}]{McKenzie_2000}
{McKenzie}, D.~E. 2000, \solphys, 195, 381

\bibitem[{{McKenzie}(2013)}]{McKenzie_2013}
---. 2013, \apj, 766, 39

\bibitem[{{McKenzie} \& {Hudson}(1999)}]{McKenzie_1999}
{McKenzie}, D.~E., \& {Hudson}, H.~S. 1999, \apjl, 519, L93

\bibitem[{{Priest}(1999)}]{Priest_1999}
{Priest}, E.~R. 1999, \apss, 264, 77

\bibitem[{{Reeves} {et~al.}(2010){Reeves}, {Linker}, {Miki{\'c}}, \&
  {Forbes}}]{Reeves_2010}
{Reeves}, K.~K., {Linker}, J.~A., {Miki{\'c}}, Z., \& {Forbes}, T.~G. 2010,
  \apj, 721, 1547

\bibitem[{{Savage} \& {McKenzie}(2011)}]{Savage_2011}
{Savage}, S.~L., \& {McKenzie}, D.~E. 2011, \apj, 730, 98

\bibitem[{{Savage} {et~al.}(2012){Savage}, {McKenzie}, \&
  {Reeves}}]{Savage_2012}
{Savage}, S.~L., {McKenzie}, D.~E., \& {Reeves}, K.~K. 2012, \apjl, 747, L40

\bibitem[{{Savage} {et~al.}(2010){Savage}, {McKenzie}, {Reeves}, {Forbes}, \&
  {Longcope}}]{Savage_2010}
{Savage}, S.~L., {McKenzie}, D.~E., {Reeves}, K.~K., {Forbes}, T.~G., \&
  {Longcope}, D.~W. 2010, \apj, 722, 329

\bibitem[{{Seaton} \& {Forbes}(2009)}]{Seaton_2009}
{Seaton}, D.~B., \& {Forbes}, T.~G. 2009, \apj, 701, 348

\bibitem[{{Tucker}(1973)}]{Tucker_1973}
{Tucker}, W.~H. 1973, \apj, 186, 285

\bibitem[{{{\v S}vestka} {et~al.}(1998){{\v S}vestka}, {F{\'a}rn{\'{\i}}k},
  {Hudson}, \& {Hick}}]{Svestka_1998}
{{\v S}vestka}, Z., {F{\'a}rn{\'{\i}}k}, F., {Hudson}, H.~S., \& {Hick}, P.
  1998, \solphys, 182, 179

\bibitem[{{Verwichte} {et~al.}(2005){Verwichte}, {Nakariakov}, \&
  {Cooper}}]{Verwichte_2005}
{Verwichte}, E., {Nakariakov}, V.~M., \& {Cooper}, F.~C. 2005, \aap, 430, L65

\bibitem[{{Warren} {et~al.}(2011){Warren}, {O'Brien}, \&
  {Sheeley}}]{Warren_2011}
{Warren}, H.~P., {O'Brien}, C.~M., \& {Sheeley}, Jr., N.~R. 2011, \apj, 742, 92

\bibitem[{{Webb} {et~al.}(2003){Webb}, {Burkepile}, {Forbes}, \&
  {Riley}}]{Webb_2003}
{Webb}, D.~F., {Burkepile}, J., {Forbes}, T.~G., \& {Riley}, P. 2003, Journal
  of Geophysical Research (Space Physics), 108, 1440

\end{thebibliography}

\end{document}